\documentclass[aps,amssymb,amsmath,prb,twocolumn]{revtex4}
 \usepackage{color}
 \usepackage{graphicx}

\begin{document}

\title{Super-Poissonian Shot Noise as a Measure of Dephasing in Closed Quantum Dots}
\author{Daria Fleurov and Eli Eisenberg}
\affiliation{Beverly and Raymond Sackler Faculty of Exact Sciences, School of Physics and
Astronomy, Tel-Aviv University, Tel Aviv 69978, Israel}

\begin{abstract}
Electron-electron interactions play a major role in determining the low-temperature rate of phase
loss of electrons in mesoscopic systems. The study of the dephasing rate is expected to
contribute to the understanding of the many-body nature of such systems. Closed quantum dots are of
special interest in this respect, due to theoretical predictions suggesting a possible transition
temperature below which the dephasing rate vanishes. This prediction has attracted much attention,
since closed quantum dots are prime candidates for storage units in quantum computers, and thus
their phase coherence properties are of great importance. However, an effective method for
measuring the dephasing rate within a closed quantum dot is still lacking. Here we
study two-level systems and show that the Fano factor has a sharp peak as a
function of the chemical potential, the location of which can be simply related to the dephasing
rate. We thus suggest to use the properties of the Fano factor peak in the super-Poissonian
regime as a probe for the dephasing rate.

\end{abstract}

\maketitle

\section{Introduction}

The study of electrons' phase coherence in mesoscopic systems has been an active field in the past
few years. One of the main contributions to the loss of quasi-particle phase coherence (or
dephasing) at low temperatures are electron-electron interactions. Thus, measurements
of the low temperatures dephasing rate provide us with a valuable probe to gauge our understanding
of the structure of the many-electron ground state. In particular, the nature of the many-body wave
function in confined geometries was predicted to change sharply at low excitation energies, where
the wave function is expected to be localized in Fock space~\cite{locinFock}. Consequently, one
expects the dephasing rate to vanish at some finite temperature, which scales like
$\Delta\sqrt{g_T/\ln g_T}$, where $\Delta$ is the mean-level spacing and~$g_T$ is the dimensionless
Thouless conductance of the system. The prediction of vanishing dephasing rate in isolated dots is
of great importance in light of the search for systems with long coherence times to be used as
storage units in quantum computers. However, it was claimed that the logarithmic correction, and
maybe even the existence of the transition are artifacts of various approximations done in the
process of mapping the many-body Hamiltonian onto the localization problem~\cite{Chaos2}. Thus, an
experimental verification of this localization transition, through measurements of the dephasing
rate in closed (i.e., nearly isolated) quantum dots, is desirable for both theoretical and
applicative reasons.

Whereas there are a number of ways to measure the dephasing times of thermalized states in open
quantum dots, the situation is much more complicated in closed dots. Folk~et~al.~\cite{FolkPatel}
have measured the magneto-conductance and used it as a probe for dephasing times. Subsequent
theoretical analysis~\cite{EisHeldPRL,EisHeldChaos} has yielded some numerical estimates of the
dephasing times, and a clear $\Delta$ dependence was observed. However, it turns out that the
magneto-conductance is not sensitive enough in the low-temperature regime. Although  the results of
Folk~et~al. are certainly consistent with vanishing dephasing rates at low temperatures, they do
not exclude other scenarios.

A renewed interest in charge and current fluctuations and their full counting statistics (FCS) has
arisen recently. These quantities provide additional insight into transport phenomena in mesoscopic
systems (See Ref.~\onlinecite{BlaBut} and references therein). For example, it has been pointed out
that the noise of the transferred charge is suppressed in non-interacting conductors and simple
quantum dots, as compared to the classical Poisson statistics~\cite{BlaBut, BagNaz}. On the other
hand, the shot noise is strongly enhanced in interacting quantum
dots~\cite{bulka,safonov,Gust-sup,cottet,kiesslich}. The full counting statistics of currents in
such mesoscopic systems thus yields a great deal of information on the underlying electronic
processes, much beyond what is revealed by the averaged current alone.

In this work we suggest to use the current fluctuations as a probe of electron dephasing in closed
quantum dots. For simplicity, we focus on a two-level model, where dephasing is manifested through
the change of the energy state of the dot. We use the noise-to-mean ratio, known as the Fano
factor~($\!F$), as a convenient measure of the shot noise. The classical shot-noise problem, with
independent and uncorrelated events, has Poisson statistics, where the Fano factor equals one. As
mentioned above, a single-level quantum dot exhibits suppressed noise, i.e.~\mbox{$F<1$}. The two
level system, on the other hand, displays super-Poissonian noise, due to the \emph{dynamical channel
blockade}~\cite{Belzig}. Here we examine how this noise enhancement depends on the dephasing rate,
i.e. the coupling between the channels.

The rest of this paper is organized as follows: in Section~\ref{sec: model} we present
the model describing the closed two-level quantum dot, and arrive at a general expression for
the Fano factor. In section~\ref{sec: zero} we proceed to discuss charge transport in the zero bias limit.
It is shown that the Fano factor exhibits a sharp peak as a function of the leads' chemical potential,
the characteristics of which can be related to the dephasing rate. In section~\ref{sec: intuitive}
we provide a qualitative description of the random transport process, and in section~\ref{sec: RMT}
we show that the effect survives ensemble averaging.

\section{Full counting statistics in a two-level quantum dot}\label{sec: model}

We describe the closed quantum dot~(QD) as a double-barrier potential well with two energy levels
in the transport window ($\varepsilon_1$ and $\varepsilon_2$), neglecting the influence of all
other energy states of the dot. Electrons can transport through the dot by means of tunneling
through the barriers. Dephasing in this isolated regime is manifested through inelastic scattering
between the two levels within the dot. An electron in the excited state, with energy
$\varepsilon_2$, can decay to the ground state, $\varepsilon_1$, with the rate $\gamma$, while an
electron in the ground state can be scattered to the excited state  with the rate
$\gamma\exp({-\delta/k_BT})$, where $\delta=\varepsilon_2-\varepsilon_1$ is the energy spacing
between the two states, $T$ is the electron temperature and $k_B$ is Boltzmann's constant. This
relation between the excitation and decay rates is dictated by detailed balance.

The dot is in the Coulomb blockade degenerate regime, where the charge on the dot can fluctuate
between $Ne$ and $(N\!+\!1\!)\,e$. Therefore, the dot can be found in one of three possible states;
$|0\rangle$ -- where the dot is empty, and $|i\rangle$, with $i=1,2$, where one of the two energy
levels is occupied. The simultaneous occupation of both energy levels is forbidden by Coulomb
blockade. The transition rates between the occupied states $|i\rangle$ and the empty state
$|0\rangle$ are
  \mbox{$\Gamma_{i\rightarrow 0} = (1\!-\!f_i^L)\,\Gamma_i^L + (1\!-\!f_i^R)\,\Gamma_i^R$} and
  \mbox{$\Gamma_{0\rightarrow i} = 2f_i^L\,\Gamma_i^L + 2f_i^R\,\Gamma_i^R$},
where $\Gamma_i^{\alpha}$ is the tunneling rate between the state $|i\rangle$ and lead~$\alpha$
(left or right), $f_i^{\alpha}$ is the Fermi occupation function in the lead,
  \mbox{$f_i^{\alpha}=1/(1+\exp((\varepsilon_i-\mu_{\alpha})/\mbox{kT}))$},
and $\mu_{\alpha}$ is the chemical potential in the lead. The factor~$2$ accounts for
the two spin states in each energy level in the dot.

Within the sequential tunneling limit, electron dynamics can be described through the master
equation formulation. In order to find the full counting statistics, we follow Bagrets and
Nazarov~\cite{BagNaz} and add counting fields $\exp(i\chi)$ to the rates involving the change of
the dot's excess charge from $0$ to $1e$ (``incoming'' events). The cumulant generating function is
then proportional to
the smallest eigenvalue of the modified rates matrix, 
 \mbox{$S(\chi)=-t_0\Lambda_{min}(\chi)$}, as long as $t_0$, the duration of a single experiment,
is longer then the average dwell time in the dot. 
The cumulants are given by
  $C_n=(-i)^n\partial^n S(\chi)/\partial\chi^n$.
Applying this approach to the general two level case, one obtains the Fano factor

\begin{widetext}
\begin{eqnarray}\label{eq: Fano, full}
    F   &=& 1 + 2\,\Bigg\{
        \frac{\Gamma_{0\rightarrow1}\Gamma_{1\rightarrow0}+
            \Gamma_{0\rightarrow2}\Gamma_{2\rightarrow0}}
        {\Gamma_{0\rightarrow1}\Gamma_{2\rightarrow0}+\Gamma_{0\rightarrow2}\Gamma_{1\rightarrow0}+
            \Gamma_{1\rightarrow0}\Gamma_{2\rightarrow0} +
            \gamma\Gamma_{in}(1+e^{-\delta/\rm{kT}}) +
            \gamma(\Gamma_{1\rightarrow0} + e^{-\delta/\rm{kT}}\Gamma_{2\rightarrow0})}
            \Bigg. \\
     & & - \Bigg.\frac{\Gamma_{in}(\Gamma_{1\rightarrow0}\Gamma_{2\rightarrow0}+
        \gamma(\Gamma_{1\rightarrow0} + e^{-\delta/\rm{kT}}\Gamma_{2\rightarrow0}))
            (\Gamma_{tot}+\gamma(1 + e^{-\delta/\rm{kT}}))}
            {\left(\Gamma_{0\rightarrow1}\Gamma_{2\rightarrow0}+\Gamma_{0\rightarrow2}\Gamma_{1\rightarrow0}+
            \Gamma_{1\rightarrow0}\Gamma_{2\rightarrow0} +
            \gamma\Gamma_{in}(1+e^{-\delta/\rm{kT}}) +
            \gamma(\Gamma_{1\rightarrow0} + e^{-\delta/\rm{kT}}\Gamma_{2\rightarrow0})\right)^2}
        \Bigg\},\nonumber
\end{eqnarray}

\begin{equation}
     {\rm where}\quad\Gamma_{in} = \Gamma_{0\rightarrow1}+\Gamma_{0\rightarrow2}\
     \qquad
     {\rm and}\quad  \Gamma_{tot} = \Gamma_{0\rightarrow1}+\Gamma_{1\rightarrow0}+\Gamma_{0\rightarrow2}
                    +\Gamma_{2\rightarrow0}.\nonumber
\end{equation}

\end{widetext}
The Fano factor does depend on the dephasing rate $\gamma$, so in principle it can be used to
determine this rate experimentally, provided all the other parameters are known. However, the
dependence on the various parameters is much too complex to allow for a practical use of this
formula. In the next section we focus on the zero bias regime, and present a setup where the
expression for $F$ can be greatly simplified.

\section{The zero-bias regime}\label{sec: zero}

Within the zero bias limit, the average current through the dot vanishes. However, electrons do
transport between the island and the leads due to charge fluctuations on the dot, and examination
of the ``entries'' statistics does yield non-trivial information. The tunneling rates now take the
form
\begin{gather}\label{eq: external rates , eV=0}
  \Gamma_{i\rightarrow 0} = (1\!-\!f_i)\,(\Gamma_i^L + \Gamma_i^R) \equiv (1\!-\!f_i)\,\Gamma_i,
      \nonumber\\
  \Gamma_{0\rightarrow i} = 2\,f_i\,(\Gamma_i^L + \Gamma_i^R)\equiv 2\,f_i\,\Gamma_i.
\end{gather}
Alternatively, one can study the setup where one of the contacts is pinched off,
leaving the dot in contact with a single lead, as illustrated in fig.~\ref{FIG: time-trace}(a,b).

In Fig.~\ref{FIG: Fano, eV=0} the zero-bias Fano factor~(\ref{eq: Fano, full}) is plotted as a
function of the chemical potential in the low temperature limit, $\rm{k_BT}\ll\delta$. Two distinct
regimes are observed. When the chemical potential in the leads,~$\mu$, is within the range of
$\rm{k_BT}$ near the ground level, noise is suppressed below the Poisson value, in agreement with
previous observations and predictions in single-level dots~\cite{Gust-sub,BagNaz}. In this case the
excited level doesn't effect the transport, as its energy is far above the chemical potential.
Accordingly, no dephasing effects are observed. However, when both energy levels are below the
chemical potential, $\mu\sim\varepsilon_2$, and the two transport channels are distinguishable
(i.e., the tunneling rates through them are different), the role of dephasing is revealed.

For $\mu$ in the vicinity of the excited level, strong noise enhancement is observed at low
dephasing rates, $\gamma\!\ll\Gamma_1$, and a peak appears in the Fano factor as a function of
chemical potential. The height of the peak, as well as its location, clearly depend on the
dephasing rate, as seen in fig.~\ref{FIG: Fano, eV=0}(a). As dephasing grows stronger the noise is
suppressed, reaching the limit of $F\sim1$ when dephasing rates are very high. Strong coupling
between the two states causes them to become indistinguishable, so that the system becomes
effectively a single energy level. In fig.~\ref{FIG: Fano, eV=0}(b) the role of the ratio between
the two tunneling rates $\Gamma_1$ and $\Gamma_2$ is demonstrated; the noise enhancement is
stronger when the excited level has stronger coupling with the leads. As this coupling becomes
weaker, the influence of the second level can be neglected, leading to the Poissonian noise of a
single level dot.

\begin{figure}
  \centering
  \includegraphics[width=90mm,keepaspectratio,clip]{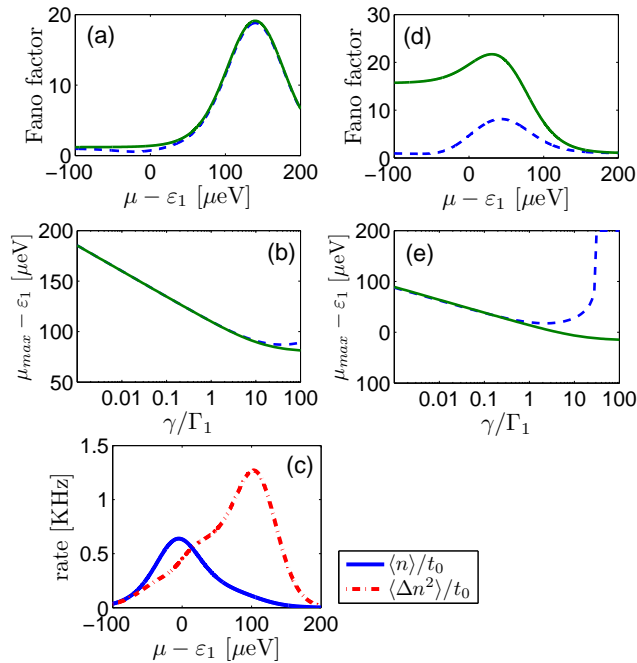}\\
  \caption{
    {A sample calculation of the cumulants $C_1$ and $C_2$, using experimentally relevant
    parameters~\cite{Gust-sup}.
    Tunneling widths are $\Gamma_1=1.6 \rm{KHz}$ and $\Gamma_2=23 \rm{KHz}$.
    The electron temperature is $T_{el}=230\rm{mK}$ (${\rm k_BT}\sim22\mu\rm{eV}$).
    Panels~(a)~to~(c) present the low temperature limit, $\delta=5kT$.
    We set the energy of the lower state $\varepsilon_1$ as zero,
    while the excited state has energy $\varepsilon_2=110\mu{\rm eV}$.
    {\bf(a)} The Fano factor, evaluated from the general solution~(\ref{eq: Fano, full}),
    compared with the zero temperature approximation~(\ref{eq: F, f1=1}), dashed and solid lines respectively.
    {\bf(b)} Peak's location for various dephasing rate values.
    The dashed line was obtained by an explicit calculation
    of the full expression~(\ref{eq: Fano, full}) at different values of $\gamma$.
    The solid line is the expression~(\ref{eq: Efmax}).
    {\bf(c)} The first two cumulants. For the conditions above and decay rate is $\gamma\sim0.09\rm{KHz}$,
    the peak is at $\mu\sim1.8\mu$eV, where the cumulants are $C_1\sim30\rm{Hz}$ and $C_2\sim600\rm{Hz}$.
    The Fano factor~{\bf(d)} and the location of peak~{\bf(e)} are calculated at a higher
    temperature to energy spacing ratio, $\delta=14\mu{\rm eV}$ ($\varepsilon_1=0$ and $\varepsilon_2=14\mu$eV).
    Note the logarithmic behaviour of the peak's location.}
  }\label{FIG: C1 C2 gustavsson}
\end{figure}

Following these findings, we now look for a more convenient expression for the Fano factor, in
order to characterize the dependence of the peak's location on the parameter $\gamma$.
Expanding~(\ref{eq: Fano, full}) to the leading order in $e^{-\delta/\rm{kT}}$
in the super-Poissonian regime, where $\mu\sim\varepsilon_2$, leads to
\begin{equation}\label{eq: F, f1=1}
    F = 1+2\,\frac{\Gamma_{2\leftarrow0}\Gamma_{0\leftarrow2}}
        {\Gamma_{1\leftarrow0}\Gamma_{0\leftarrow2}+\gamma(\Gamma_{1\leftarrow0}+\Gamma_{2\leftarrow0})}.
\end{equation}
In order to determine the location of the peak, $\mu_{\max}$, we use the fact that the Fermi
occupation function $f_1$ is almost constant in the super-Poissonian regime, and so the line shape
of $F$ is determined predominantly by the change of the function $f_2$. Using this approximation,
one obtains
\begin{equation}\label{eq: Efmax}
    \mu_{\max}-\varepsilon_2 =
    -\frac{1}{2}\,\rm{k_BT}\,\log\left(\frac{1/\Gamma_1+1/\Gamma_2}{1/\gamma+1/\Gamma_2}\right).
\end{equation}
When $(\Gamma_1,\gamma)\ll\Gamma_2$ a further simplification is possible,
\begin{equation}
    \mu_{\max}-\varepsilon_2 = -\frac{1}{2}\,\rm{k_BT}\,\log(\gamma/\Gamma_1).
\end{equation}
Figure~\ref{FIG: C1 C2 gustavsson} demonstrates the quality  of this zero temperature approximation
for both the Fano factor and the peak's location. It should be noted that the logarithmic
dependence of the peaks' location on $\gamma$ persists for higher temperatures (figure ~\ref{FIG:
C1 C2 gustavsson}(e)), where the approximation~(\ref{eq: F, f1=1}) breaks down.

\begin{figure}
  \centering
  \includegraphics[width=90mm,keepaspectratio,clip]{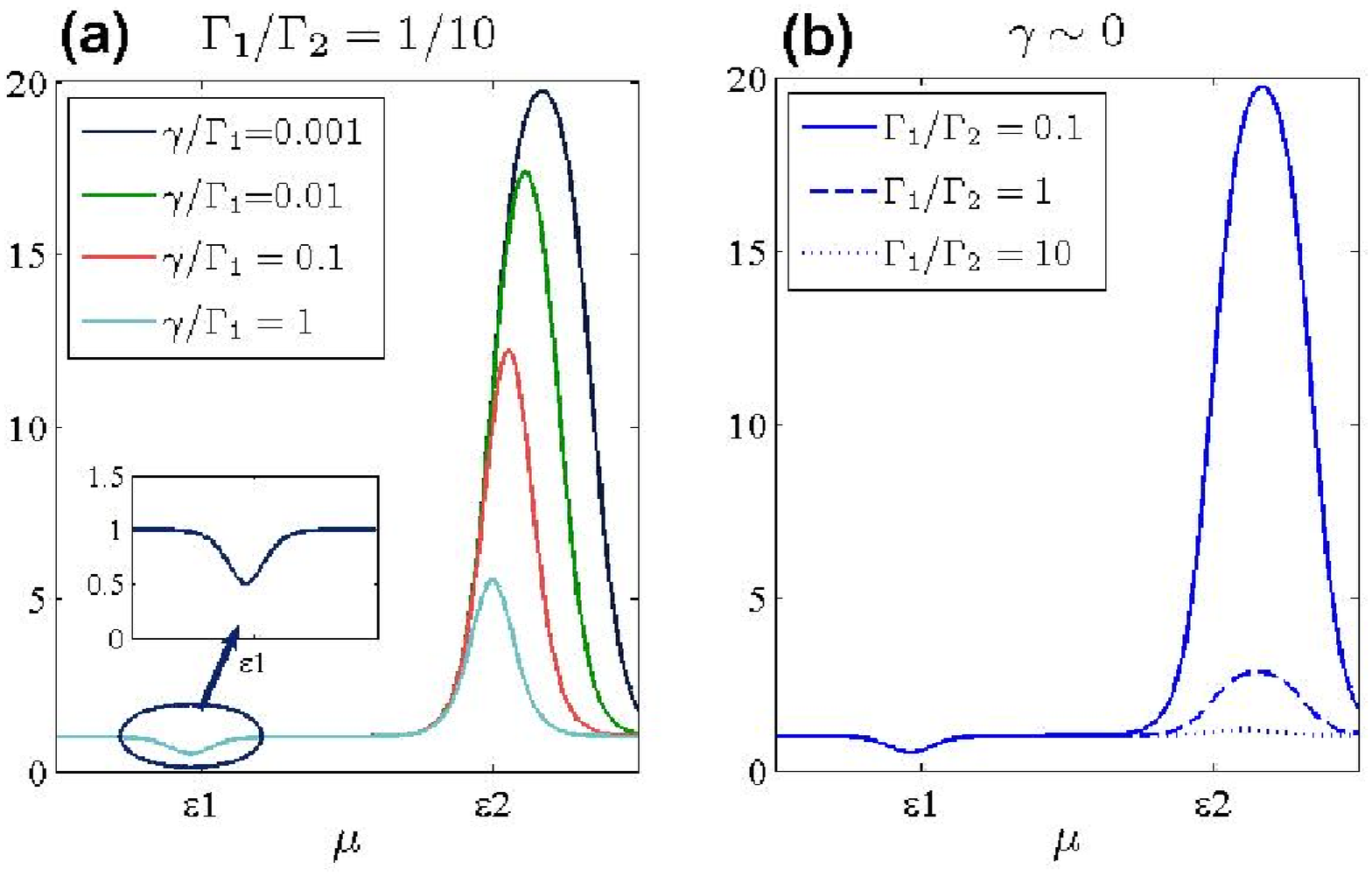}\\
  \caption{
    \textbf{(a)} The Fano factor is plotted for several values of dephasing rates at
            a very low temperature (${\rm k_B T}/\delta=1/20$),
      In the super-Poissonian region the Fano factor shows a clear peak, the height of which
       depends on dephasing.
      Note the shift in the location of the peak with the changing of $\gamma$.
    {Inset:} $F$ reaches a minimum value of $1/2$, when $\Gamma_{1\leftarrow0}=\Gamma_{0\leftarrow1}$.
    \textbf{(b)} The Fano factor for various values of the ratio $\Gamma_1/\Gamma_2$, and very week dephasing.
      $F$ is dramatically increased when the coupling of the ground level $\varepsilon_1$
      with the lead is considerably weaker then that of the excited level $\varepsilon_2$.
  }\label{FIG: Fano, eV=0}
\end{figure}

\section{Qualitative description}\label{sec: intuitive}

\begin{figure}
  \includegraphics[width=80mm]{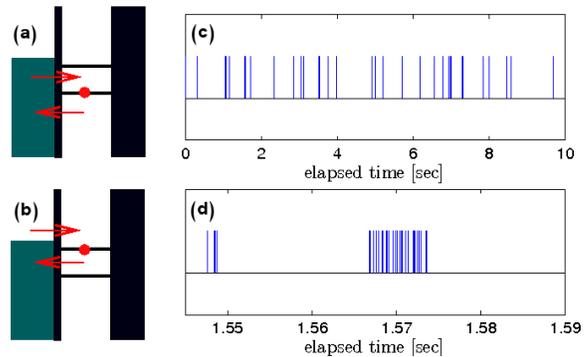}\\
  \caption{
     Transitions in a single lead dot.
     (a) a ``slow'' process, involving the occupation of the ground level.
     (b) a ``fast'' process, involving the excited state.
     (c)~and~(d) present typical time traces, obtained from a numerical
     experiment, with parameters similar to those in fig.~\ref{FIG: C1 C2 gustavsson}. In (c) the
     entire experiment is shown, where the stems are registered events. In the finer resolution in
     (d) the fast processes can be seen.
  }\label{FIG: time-trace}
\end{figure}

The super-Poissonian behavior in the zero-bias regime can be intuitively understood within the
dynamical channel blockade picture of Ref.~\onlinecite{Belzig}. The super-Poissonian statistics is
most pronounced at low temperatures for $\Gamma_1\!\ll\Gamma_2$ and $\gamma\sim0$, as seen in
fig.~\ref{FIG: Fano, eV=0}. In this case the Fano factor reaches its maximum value at
$\mu_{\rm{max}}$ {\it above} the excited resonant level $\varepsilon_2$, and the transport is
dominated by the two processes illustrated in~\ref{FIG: time-trace}(a,b), where an empty dot is
occupied quickly, entering one of the two allowed states. Due to the low coupling between the
states, an electron is most likely to remain in the same state until it exits the dot. The entry
rate is considerably faster then the exit rate, so one can ignore the time it takes an electron to
enter the dot, and thus in both processes (a) and (b) the time intervals between entries are
distributed exponentially. Most of the states in the lead at the energy $\varepsilon_1$ are
occupied, and therefore an electron in the ground level has a lower exit rate then one in the
excited state. Thus the process is a combination of two exponential processes, one of them faster
then the other.

Within this picture, the time evolution of the dot's occupation is as follows: an empty dot is
occupied immediately. If the electron enters the ground state, transport is blocked for the longer
time scale $1/\Gamma_{0\leftarrow1}$ until the electron leaves the dot (process~(a)). If the
excited state is entered, transport is blocked for a much shorter period (process~(b)).
A series of such ``fast'' transitions would seem as a bunch of electrons tunneling into the
dot and out of it at practically the same time.
    Each time an electron leaves the dot is registered as an ``event'' on the time-trace plotted
in~~fig.~\ref{FIG: time-trace}(c,d). These figures clearly show how these events are clustered into
groups of random size, which are separated by exponentially distributed time intervals of the
longer time scale $1/\Gamma_{0\leftarrow1}$. The individual fast events within a cluster are
separated by the much shorter time scale~$1/\Gamma_{0\leftarrow2}$. Each cluster is terminated by
an electron entering the ground state.

    The resulting statistics display positive correlation between events, giving rise to a
super-Poissonian distribution. Now the role of dephasing can be easily interpreted: inelastic scattering
events bring about a finite probability for an excited dot to decay into the ground state, thus
increasing the probability of terminating a cycle of fast events. As a result, the
clusters become shorter and the correlation weaker. At high dephasing rates the transport is
effectively governed by a single exponential process, just like the classical shot-noise, resulting in
a Poissonian distribution.

Following the above description, one looks at two transport channels with different
``conductances'', one significantly larger then the other. It is possible to derive the full
counting statistics of the transport process in this case, which leads to the Fano factor,
  \begin{equation}\label{eq: Fano, p}
    F=1+\frac{2\,p}{{1-p}},
    \end{equation}
where $p$ is the probability to tunnel through the open channel~\cite{Belzig}. In a QD with a
single contact and $\Gamma_1\!\ll\Gamma_2$, the open channel corresponds to the fastest countable process
in fig.~\ref{FIG: time-trace}(b), and its probability is given by
\begin{equation}\label{eq: prob fast event}
  p = \frac{\Gamma_{2\leftarrow0}}{\Gamma_{1\leftarrow0}+\Gamma_{2\leftarrow0}}
        \times
      \frac{\Gamma_{0\leftarrow2}}{(\Gamma_{0\leftarrow2}+\gamma)}.
\end{equation}
Plugging this into the expression for the Fano factor~(\ref{eq: Fano, p}), one arrives
at~(\ref{eq: F, f1=1}). Note however, that this reasoning can be applied at low temperatures and low
dephasing rates only, where it is reasonable to assume exponentially distributed processes in both
channels.  When either temperature or dephasing are higher a sequence of several transitions could
occur between two subsequent counts, such as
    $|0\rangle\rightarrow|2\rangle\rightarrow|1\rangle\rightarrow|0\rangle\rightarrow|2\rangle$,
where only the two transitions that involve $|0\rangle$ as the initial state are registered. The
time intervals in this scenario have a much more complicated distribution.

\section{Ensemble averaging of the Fano factor}\label{sec: RMT}

In the previous sections, a dephasing-sensitive super-Poissonian noise peak was demonstrated for
the two-level QD with a single contact, where particular parameters where used, such as
$\Gamma_1\!\ll\Gamma_2$ and temperature was low. In an attempt to attack the more general setup, we
show here that this effect is significant not only in some specifically chosen systems, but also for
the ensemble averaged Fano factor.  Within random matrix theory~\cite{RMT},
the tunneling rates are random variables with the Porter-Thomas distribution.
\begin{equation}\label{eq: Port-Thomas}
    P(\Gamma) = \left\{
        \begin{array}{ccc}
            \frac{1}{\sqrt{2\pi \bar{\Gamma}\Gamma}}
                e^{-\Gamma/2\bar{\Gamma}}
                & \mbox{(GOE)} \\
        1/\bar{\Gamma}
                e^{-\Gamma/\bar{\Gamma}}
                & \mbox{(GUE)}
        \end{array}
        \right. ,
\end{equation}
where the tunneling rates through a given barrier average to the value $\bar\Gamma$ regardless of
energy.

The inelastic scattering rate between two states
is proportional to the squared matrix element of the interaction $V$,
\begin{equation}\label{RMT: bracket V}
\gamma_{ij}  \propto  \left| \langle i|V|j\rangle \right|^2,
\end{equation}
where  $\psi_i=|i\rangle$ is an eigenstate of $H_0$. Whenever random-matrix theory is applicable,
the spatial correlations of the electron
wave function decay over a range of order of the Fermi wavelength $\lambda_F$~\cite{RMT},
which is much smaller than the typical size of the dot ($\lambda_F\ll L$). Assuming the interaction
potential~$V$ varies slowly over a scale of $\lambda_F$, the integration
 in~(\ref{RMT: bracket V}) can be approximated by a sum over small volume elements of linear size $\lambda_F$,
\begin{equation}\label{RMT: int Vij}
    \int_{\Omega} d\vec{r}\, V(\vec{r})\psi^*_i \psi_j  \sim
    \sum_{\Omega_k} V(\vec{r}_k)\psi^*_i(\vec{r}_k) \psi_j(\vec{r}_k) .
\end{equation}
where $V(\vec{r}_k)$ is constant within each volume element $\Omega_k$.
The wave functions are distributed randomly, and the dot's linear size is typically much larger
than $\lambda_F$. Therefore, the matrix element, given by
the weighted sum of many
product terms $\psi_i^*(\vec{r}_k)\psi_j(\vec{r}_k)$, is normally distributed. Accordingly, the
dephasing rate
$\gamma$ is proportional to the square of this matrix element, hence it is also distributed
according to the Porter-Thomas distribution with an
average value of $\overline{\gamma}$.

\begin{figure}
    \includegraphics[width=90mm]{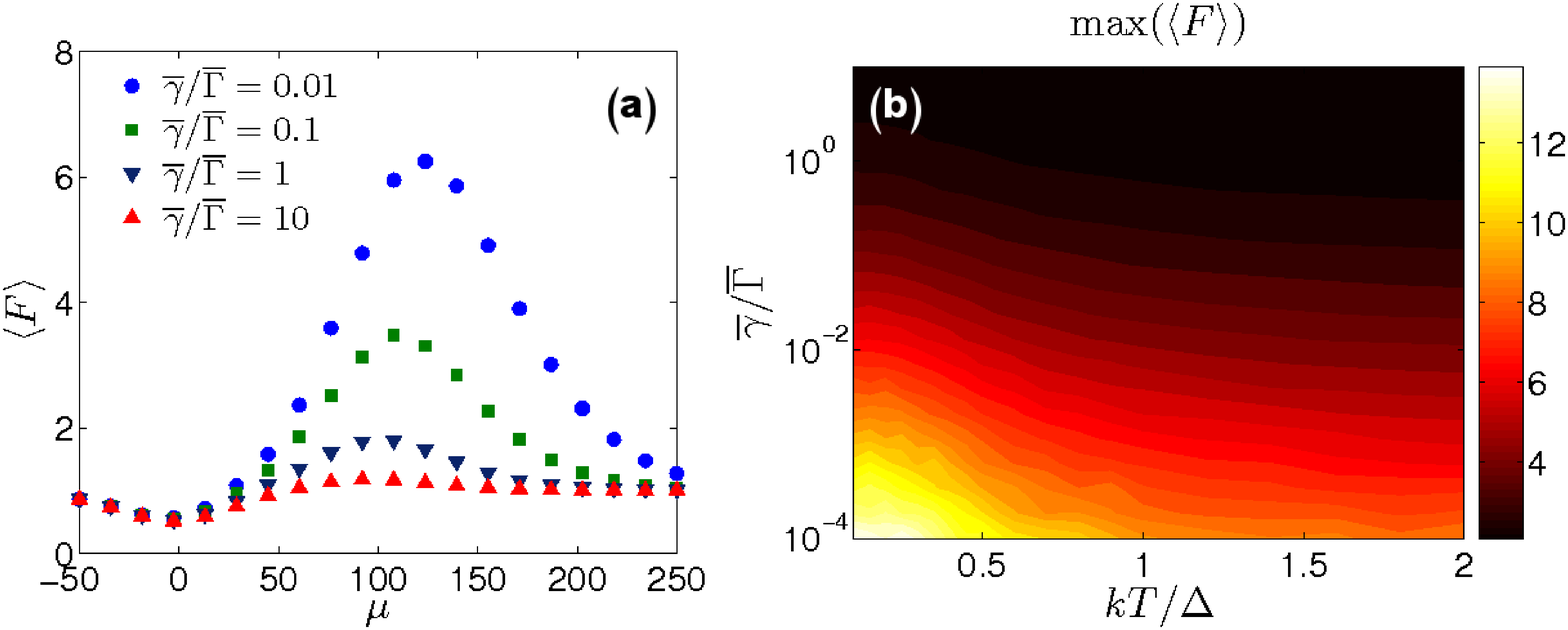}\\
    \caption{
        {\bf(a)} The averaged Fano factor as a function of the leads' chemical potential at a low
        temperature. A peak in the noise to mean ratio is observed. Here $\varepsilon_1=0$,
        $\varepsilon_2=100$ and ${\rm k_B T}=20$ (arbitrary units).
        \newline {\bf(b)}~The peak height for varying
        values of $\bar\gamma/\bar\Gamma$ and $kT/\Delta$.
        Strong enhancement is observed when average dephasing rate is much lower then the average
        tunneling rate, and temperature is lower then the energy spacing. The Poisson value $\langle\!F\rangle=1$
        is reproduced for either large dephasing rate or high temperature.}\label{FIG: aveF}
\end{figure}

Figure~\ref{FIG: aveF} presents the results of the GUE ensemble averaging over $\Gamma_1,\Gamma_2$
and $\gamma$. The averaged Fano factor is plotted as a function of the chemical potential, and,
similarly to the particular cases discussed above, exhibits a clear peak.  The dependence of the
peak's height on $\bar\gamma/\bar\Gamma$ and ${\rm k_BT}/\Delta$ is also given. The peak height
decreases with temperatures and dephasing rates, but should be clearly seen for ${\rm
k_BT}\sim\Delta$ and low dephasing rates. Thus, the Fano factor seems to be a proper probe for
observing the phase transition predicted in ref.~\onlinecite{locinFock}.

In summary, we study the peak in the Fano factor in the super-Poissonian regime and its
dependence on the inelastic scattering rate and other relevant parameters. A convenient and simple
analytical expression is developed for the peak location. Using experimentally relevant parameters,
we estimate that the effect can be easily observed, as the average rate of events at the Fano factor peak
($\mu\sim140\mu\rm{eV}$) is of the order of $30$Hz (see figure~\ref{FIG: C1 C2 gustavsson}(c)),
where experiments of this
kind were reported to last for as long as several minutes\cite{Gust-sup}.
We thus propose to use the dephasing
rate dependence as a probe for dephasing rates in closed quantum dots. This will enable
measurements which are of great importance for any likely use of quantum dots as quantum computer
memory units. The effect is still pronounced even after ensemble
averaging. Further work is needed to extend the above analysis to the general multi-level dot.

\begin{acknowledgments}{E.E. acknowledges support from an Alon fellowship at Tel-Aviv
University.}\end{acknowledgments}

\end{document}